\documentclass{article}

\usepackage{lineno,hyperref,cite}
\usepackage{graphicx,amsmath,caption,epstopdf,authblk}
\usepackage[utf8]{inputenc}
\usepackage[english]{babel}
\usepackage{amsthm,amsmath}

\title{The human monogamy behavior can influence the transmission of AIDS}
\author[a]{Chentong Li\thanks{Corresponding author: lichentong1989@stu.xjtu.edu.cn}}
\author[b]{Jiawei Liu}
\author[a]{Yicang Zhou}
\affil[a]{School of Mathematics and Statistics, Xi'an Jiaotong University, Xi'an, 710049, P.R.China}
\affil[b]{Department of Ecology and Evolution, University of Chicago, Chicago, IL 60637, USA}
\begin{document}

\maketitle

\begin{abstract}
In this letter, we mainly consider an MSM (men have sex with men) 
network to analysis how monogamy behavior can influence the transmission of HIV.
By calculating and analyzing the basic reproductive number of that network, we find the condition for when the monogamy rate
can have a positive influence on controlling the transmission of HIV. Numerical simulations are also done to illustrate that monogamy can 
influence the transmission process of HIV.
\end{abstract}

\section{Introduction}

In the transmission of AIDS, the new infection transmitted between homosexual partners, 
especially between male homosexuals, is one of the main routes of transmission of HIV\cite{grant2010preexposure,fox2001gonorrhea}. 
In recent years, with the development of the diversity of social methods, the average number of partners of a male homosexual has been
significantly  increased, resulting in the wider spread of HIV in some way\cite{grosskopf2014use,muessig2013putting}.
Meanwhile, since Netherlands becoming the first country that passed the law allowing marriage between homosexuals, 28 countries
have passed this law up to 2019. At the same time, in many other regions, the homosexual population are more and more inclining to keep
closer relationships with their sole partners\cite{wikisamesex}. In order to analyze how these new changes can influence
the transmission of sexual transmitted diseases in the homosexual population, new models need to be constructed.

Complex network and the disease transmission models based on network structure are excellent tools in 
the research of how disease transmission can be influenced by the social structure, and they have gained many 
great achievements in the field of disease transmission, forecast and 
control\cite{xiao2011modelling,peng2013vaccination,ruan2012epidemic}. 
In the earliest work of Newman et. al\cite{newman2002spread}, the threshold values of variables
about common and sexual transmitted diseases in the network structure were given. Those achievements were 
further generalized by Volz et. al\cite{volz2008sir}, and the ordinary differential equation model for the 
disease transmitted in the network structure was given. The work of Ruan et. al\cite{ruan2012epidemic} 
tells how media information could influence human behaviors, and then decrease the disease transmission rate. 
Without any exception, these work extends people's knowledge about infectious diseases, 
and this letter is based on these work.

In this letter, we first introduce the basic reproductive number\cite{dietz1993estimation},
which is a threshold to determine whether the disease will occur or not,
of the sexual transmitted disease in the network and analyze how the
monogamy rate can influence that number. Then the simulations are done
to illustrate our results.

\section{Basic reproductive number}
For an arbitrarily static network with the degree distribution $p(k)$, the
probability of an random selected node with degree $k$ is
$p_n(k)=kp(k)/\langle k \rangle$.
when there is a person selected to be infected by an epidemic in the network,
the basic reproductive number of this epidemic is \cite{miller2011edge}
$$
R_0=\sum_k p_n(k)(k-1)\frac{\beta}{\beta+\gamma}
=\frac{\beta}{\beta+\gamma}\frac{\langle k^2-k\rangle}{\langle k \rangle},
$$
where $\beta$ is the infection rate and $\gamma$ is the recovery rate.

Assuming that at the beginning of the epidemic, the fraction of people tending to
protect themselves from this epidemic (such as giving themselves vaccines) is $a$ and the
effective rate of this protection is $b$. Then the basic
reproductive number of this epidemic transmitted in the static network with some protected
people is
$$
R_0^a=\sum_k p_n(k)(k-1)(ab+(1-a))\frac{\beta}{\beta+\gamma}
=(ab+1-a)\frac{\beta}{\beta+\gamma}\frac{\langle k^2-k\rangle}{\langle k \rangle}.
$$
As $0<a<1$ and $0<b<1$, it is easy to conclude that $a(1-b)>0$, so $R_0>R_0^a$.
That means even a little awareness of people to protect themselves from the epidemic can reduce the
transmission ability of the epidemic in the network.

For a sexual transmitted disease which is transmitted in the network, we can consider a
bi-partial network, in which the vertical set can be decomposed into two disjoint sets
such that no two vertices within the same set are adjacent. Assuming one of the
vertices set is named as vertical set $1$ and another is $2$, and in the whole vertical set, the proportion of vertices belonging to
vertical set $1$  is $p$, so the the proportion of vertices belonging to vertical set $2$ is $1-p$.
We also assume that the infection rate which describe vertical set $1$ infecting $2$ is $\beta_1$ and
that describing $2$ infecting $1$ is $\beta_2$. Hence the basic reproductive number of this network is
\begin{align*}
R_0 &=p\sum_k p_n(k)(k-1)\frac{\beta_1}{\beta_1+\gamma}+(1-p)\sum_k
p_n(k)(k-1)\frac{\beta_2}{\beta_2+\gamma}  \\
&= (p\frac{\beta_1}{\beta_1+\gamma}+(1-p)\frac{\beta_2}{\beta_2+\gamma})
\frac{\langle k^2-k\rangle}{\langle k \rangle}.
\end{align*}

Then following the above equation we can get how the basic reproductive number varies with the monogamy rate.
Assuming when the people in vertical set $2$ is not the partner of people in vertical set $1$
the infection rate of the people in vertical set $1$ to infect the people in vertical set $2$ is
$\alpha\beta_1$ and will be $\rho\beta_1$ else. And the rate that a person in vertical set $2$ to infect
his non-partner neighbor is $\alpha\beta_2$ and will be $\rho\beta_2$ when to infect his partner.
Since the monogamy behavior can reduce the contact with other sexual friends,
we assume $0<\alpha<1$ and $\rho>1$. Hence
$$
R_0=(p\frac{\alpha\beta_1}{\alpha\beta_1+\gamma}+
(1-p)\frac{\alpha\beta_2}{\alpha\beta_2+\gamma})
\frac{\langle k^2-2k\rangle}{\langle k \rangle}+\frac{\rho\beta_1}{\rho\beta_1+\gamma}
+\frac{\rho\beta_2}{\rho\beta_2+\gamma}.
$$

Since it is impossible for everyone in the network to obey the monogamy, we
assume that the number of vertices in vertical set $1$ obeying the monogamy is $m$
and in vertical set $2$ the number is also $m$. Meanwhile we assume that the
number of vertices in vertical set $1$ is $v_1$ and in vertical set $2$ is $v_2$.
Then the basic reproductive number in such situations can be divided into
the sum of two parts, the first part which can be named as monogamy part is
\begin{align*}
R_0^m &=\frac{v_1}{v_1+v_2}\frac{m}{v_1}\sum_k
p_n(k)[(k-2)\frac{\alpha\beta_1}{\alpha\beta_1+\gamma}+
\frac{\rho\beta_1}{\rho\beta_1+\gamma}]
+ \\ & \quad \frac{v_2}{v_1+v_2}\frac{m}{v_2}\sum_k
p_n(k)[(k-2)\frac{\alpha\beta_2}{\alpha\beta_2+\gamma}+
\frac{\rho\beta_2}{\rho\beta_2+\gamma}] \\
&=\frac{m}{v_1+v_2}[\frac{\langle k^2-2k\rangle}{\langle k
\rangle}(\frac{\alpha\beta_1}{\alpha\beta_1+\gamma}+\frac{\alpha\beta_2}{\alpha\beta_2+\gamma})
+\frac{\rho\beta_1}{\rho\beta_1+\gamma}
+\frac{\rho\beta_2}{\rho\beta_2+\gamma}].
\end{align*}
The second part which can be named as the single part is
\begin{align*}
R_0^s & =\frac{v_1}{v_1+v_2}\frac{v_1-m}{v_1}\sum_k p_n(k)(k-1)\frac{\beta_1}{\beta_1+\gamma}
+\\ & \quad \frac{v_2}{v_1+v_2}\frac{v_2-m}{v_2}\sum_k p_n(k)(k-1)\frac{\beta_2}{\beta_2+\gamma}
\\&=\frac{v_1-m}{v_1+v_2}\frac{\langle k^2-k\rangle}{\langle k \rangle}
\frac{\beta_1}{\beta_1+\gamma}+\frac{v_2-m}{v_1+v_2}\frac{\langle k^2-k\rangle}
{\langle k \rangle}\frac{\beta_2}{\beta_2+\gamma}.
\end{align*}
Hence we can get the basic reproductive number which consider monogamy rate
$$
R_0=R_0^m+R_0^s.
$$

In the above statement we assume $0<\alpha<1$ and $\rho>1$ and as the function
$x/(x+c)$ is an increasing function for the variable $x$,  we can easily get that
$\beta/(\beta+\gamma)$ is greater than $\alpha\beta/(\alpha\beta+\gamma)$ and
less than $\rho\beta/(\rho\beta+\gamma)$. Which follows,
$$
Q_1=(\frac{\beta_1}{\beta_1+\gamma}+\frac{\beta_2}{\beta_2+\gamma})-
(\frac{\alpha\beta_1}{\alpha\beta_1+\gamma}+\frac{\alpha\beta_2}{\alpha\beta_2+\gamma})>0,
$$
and
$$
Q_2=(\frac{\rho\beta_1}{\rho\beta_1+\gamma}+\frac{\rho\beta_2}{\rho\beta_2+\gamma})-
(\frac{\beta_1}{\beta_1+\gamma}+\frac{\beta_2}{\beta_2+\gamma})>0.
$$
Hence the derivation of $R_0$ with respect to $m$ is
\begin{align*}
    \frac{\partial R_0}{\partial m} &=\frac{\partial R_0^m}{\partial m}+
    \frac{\partial R_0^s}{\partial m}\\
    &=\frac{1}{v_1+v_2}(-\frac{\langle k^2-2k\rangle}{\langle k \rangle}Q_1+Q_2).
\end{align*}

The formula $\frac{\partial R_0}{\partial m}$ will be greater than $0$,
if $\frac{\langle k^2\rangle}{\langle k \rangle}<Q_2/Q_1+2$.
So for a given network G, if its mean degree is very big
but its second moment of degree distribution is small, then
the increase of monogamy rate can reduce the infectious ability of that disease.

For example when we consider that the degree distribution of the network satisfies the Poisson distribution,
the formula $\langle k^2\rangle/\langle k \rangle$ becomes
$$
\frac{\langle k^2\rangle}{\langle k \rangle}=
\frac{\sum_kk^2e^{-\lambda}\frac{\lambda^k}{k!}}{\sum_kke^{-\lambda}\frac{\lambda^k}{k!}}
=\frac{\lambda^2+\lambda}{\lambda}=\lambda+1.
$$
Then the condition $\frac{\langle k^2\rangle}{\langle k \rangle}<Q_2/Q_1+2$ becomes $\lambda<Q_2/Q_1+1$.
Thus if the degree distribution of the social network satisfies the Poisson distribution and the
expectation number of neighbors is small, the increase of monogamy rate can reduce the
$R_0$.

For a scale free network whose degree distribution $p(k)$ satisfying
$p(k)\sim k^{-\theta}$, and whose parameter $\theta$ satisfying $2<\theta<3$
$$
\frac{\langle k^2\rangle}{\langle k \rangle}=
\frac{\sum_k k^{2-\theta}}{\sum_k k^{1-\theta}}.
$$
By numerical methods, we can find out that as the increase of the parameter $\theta$ the above equation will
become smaller. So that means in the scale free network, if the difference of neighbor numbers between different vertices
is big, then the increase of monogamy rate can reduce the $R_0$.

\section{simulation}
The simulations are done to illustrate how the different parameters can influence the transmission of epidemic.
We use the simulation method mentioned in \cite{miller2011edge} and
consider infection process and removal process in the network to simulate how the number of infected vertices vary with time.
The simulated network is a bi-partial network, where the vertical set $1$ is the set of insertive partners
and set $2$ represents receptive partners.
All of the source code files are available at \url{https://github.com/ChentongLi/social-network-simulation}.

\begin{figure}[htp]
\centering
\includegraphics[width=0.9\textwidth]{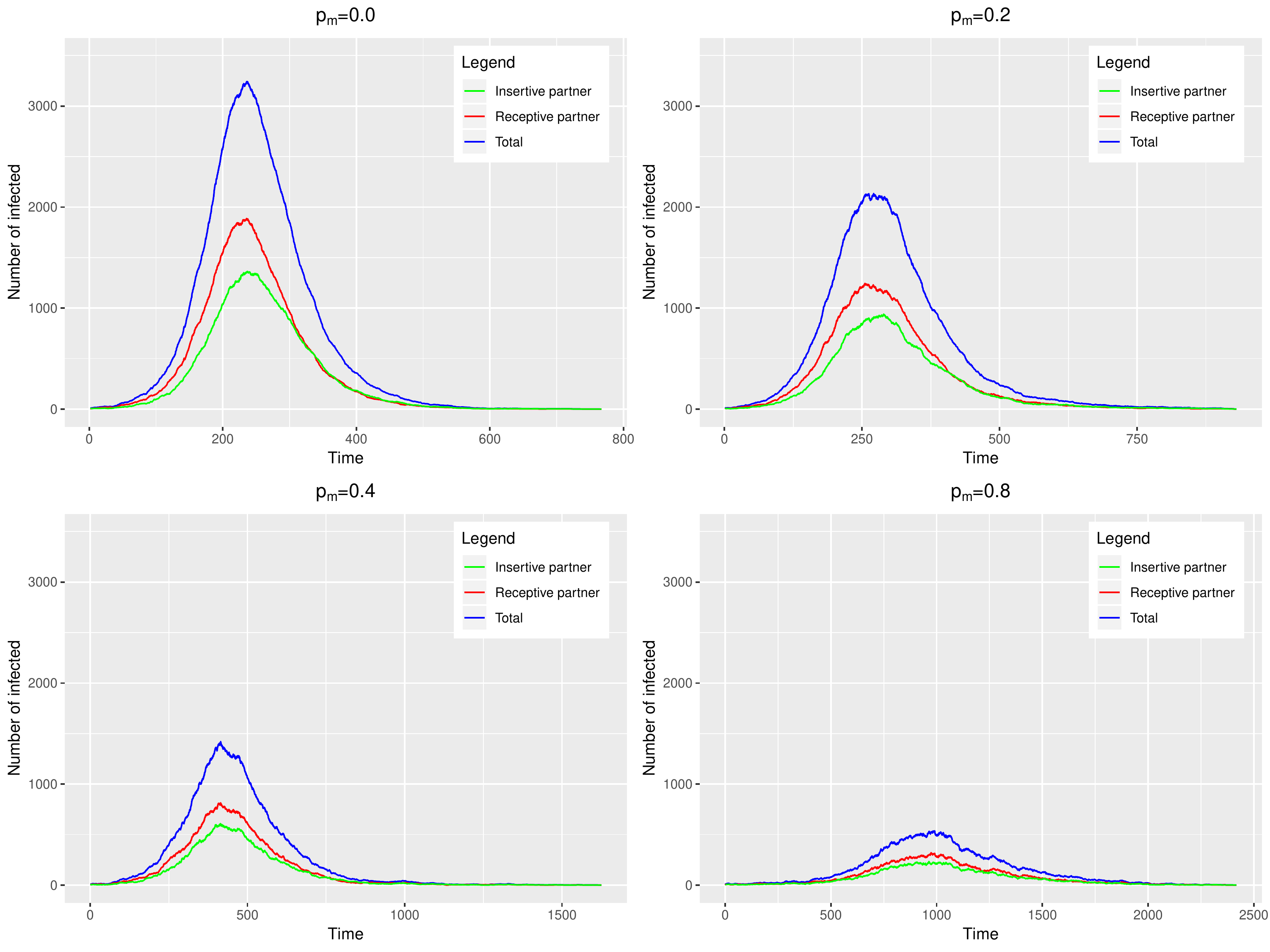}
\caption{The simulation results of number of infected vertex vary with time on the Poisson random network
with mean degree $5$ and monogamy rate (a)
$p_m=0$, (b)$p_m=0.2$, (c)$p_m=0.4$ and (d)$p_m=0.8$.}
\label{marry5}
\end{figure}

\begin{figure}[htp]
\centering
\includegraphics[width=0.9\textwidth]{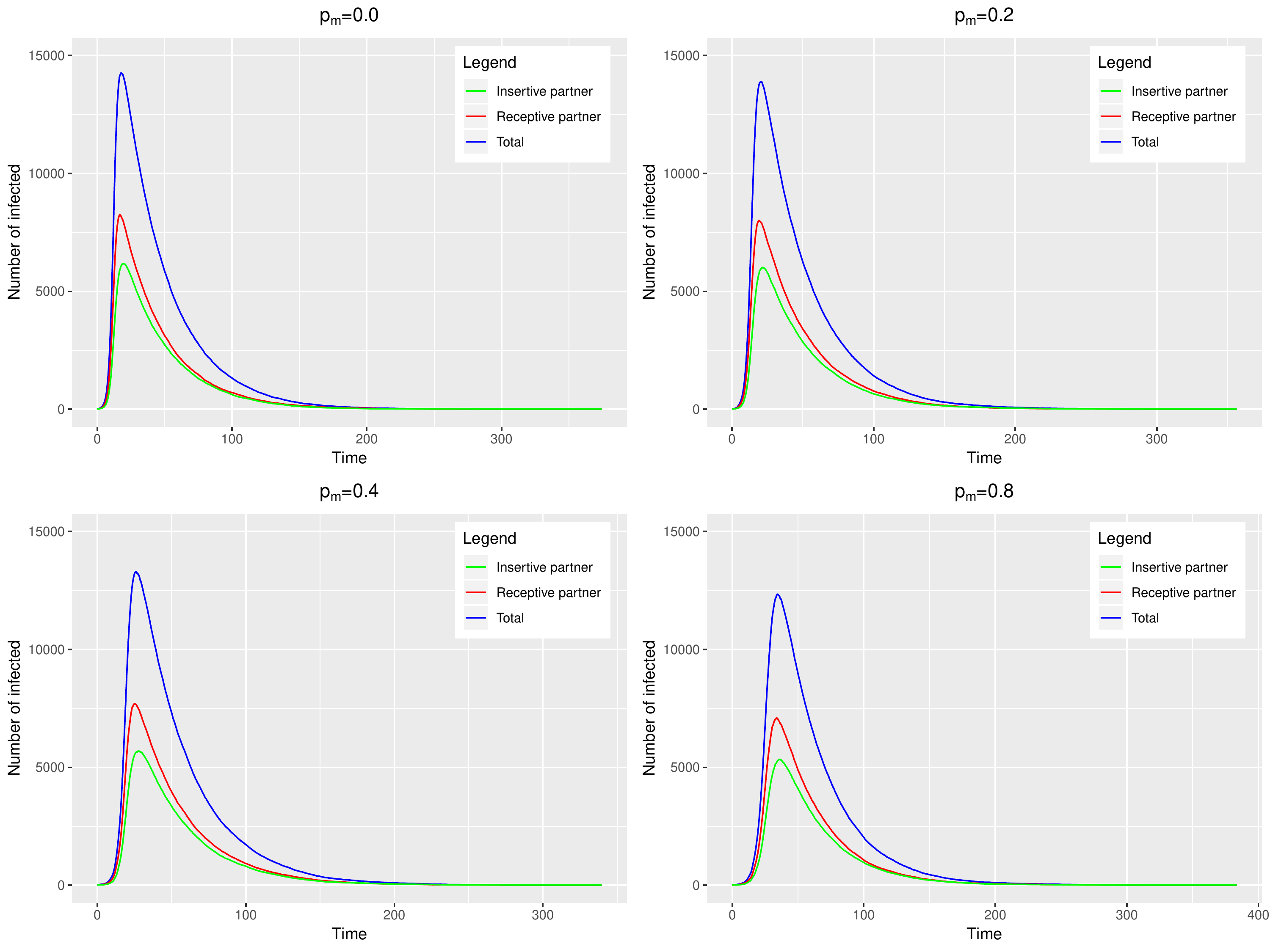}
\caption{The simulation results of number of infected vertex vary with time on the Poisson random network
with mean degree $50$ and monogamy rate (a)
$p_m=0$, (b)$p_m=0.2$, (c)$p_m=0.4$ and (d)$p_m=0.8$.}
\label{marry50}
\end{figure}

The figure.\ref{marry5} shows how the marry rate (after marriage, people will obey monogamy) $p_m=\frac{2m}{v_1+v_2}$ 
can influence the transmission of epidemic in a MSM Poisson random network with mean degree $5$ and the figure.
\ref{marry50} shows the same thing with the mean degree $50$ in the network.
These two figures illustrate that with the increase of the monogamy rate, the peak of infection process will decrease and
the maximal number of infected vertices is more sensitive to the monogamy rate in the network with a low mean degree.

\begin{figure}[htp]
\centering
\includegraphics[width=0.9\textwidth]{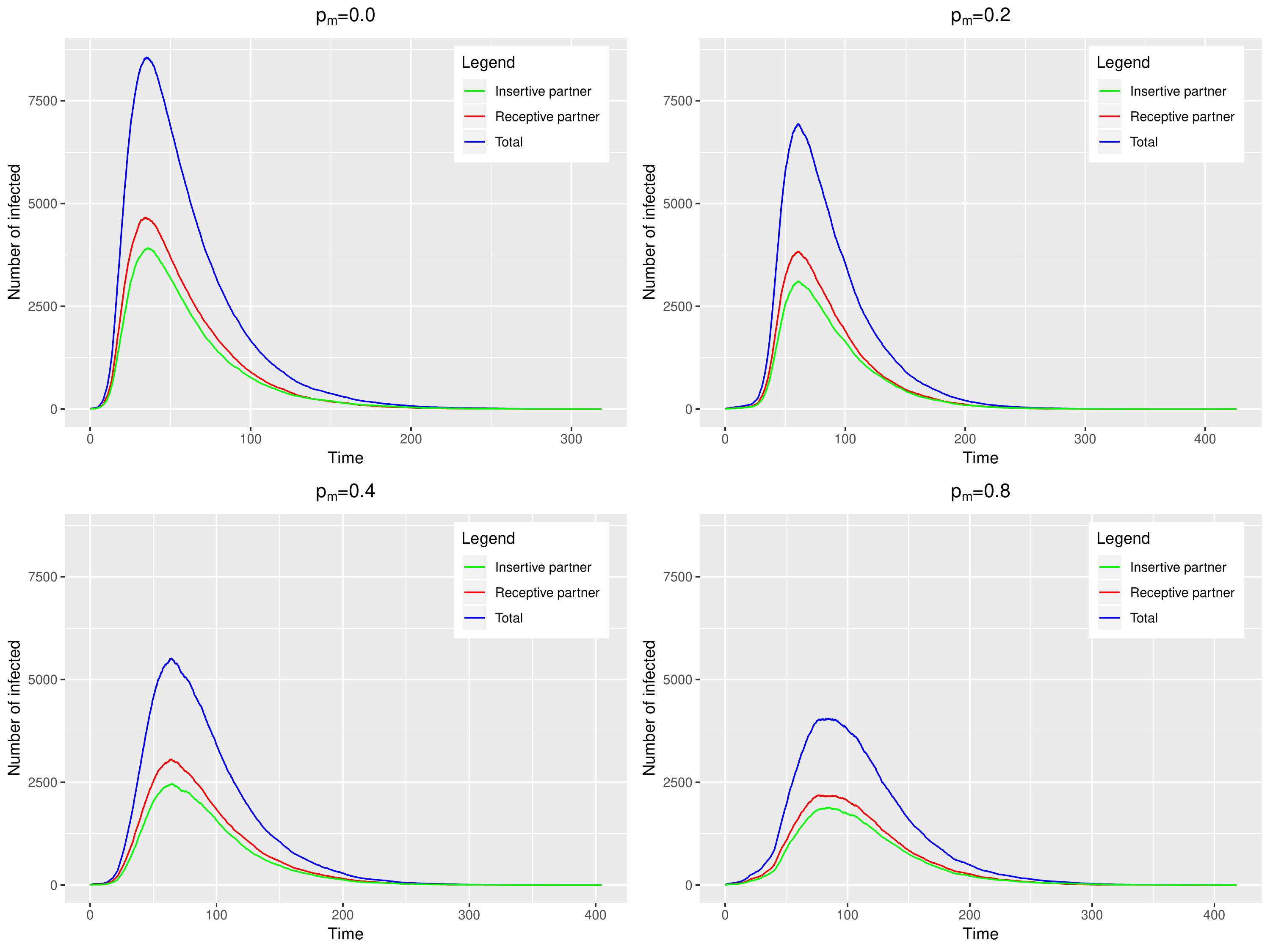}
\caption{The simulation results of number of infected vertex vary with time on the scale free random network
with the degree distribution $p(k)\sim k^{-2.1}$ and monogamy rate (a)
$p_m=0$, (b)$p_m=0.2$, (c)$p_m=0.4$ and (d)$p_m=0.8$.}
\label{marry2.1}
\end{figure}

\begin{figure}[htp]
\centering
\includegraphics[width=0.9\textwidth]{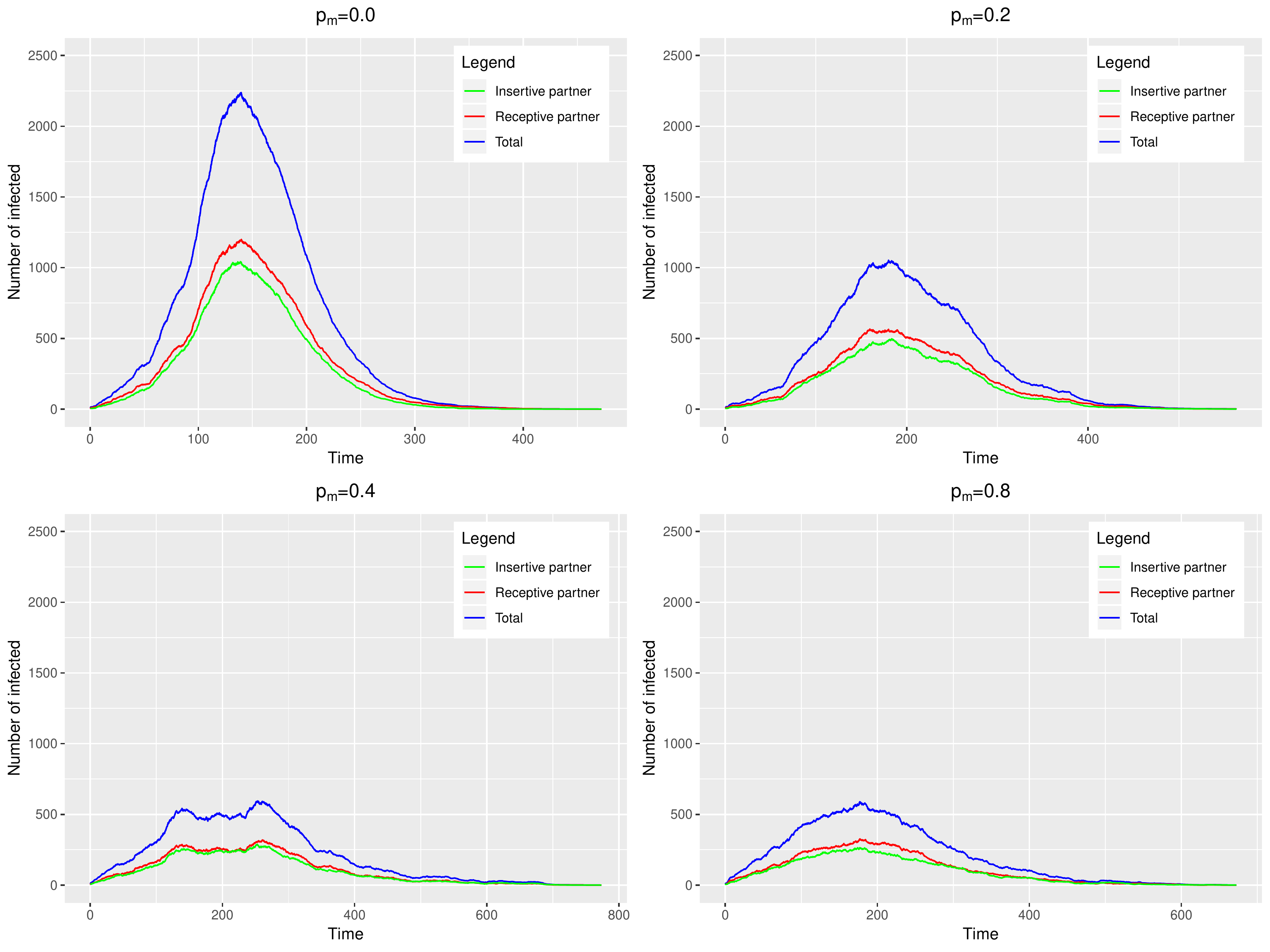}
\caption{The simulation results of number of infected vertex vary with time on the scale free random network
with the degree distribution $p(k)\sim k^{-2.99}$ and monogamy rate (a)
$p_m=0$, (b)$p_m=0.2$, (c)$p_m=0.4$ and (d)$p_m=0.8$.}
\label{marry2.99}
\end{figure}

The figure.\ref{marry2.1} and \ref{marry2.99} illustrate the simulation results of epidemic transmission on the
scale free network with the degree distribution $p(k)\sim k^{-2.1}$ and $p(k)\sim k^{-2.99}$. These two figures
show that with the increase of $\theta$, the peak of infection process will decrease and the higher monogamy rate
can make a lower maximal number of infected vertices.



\section{discussion}

In this letter, we mainly report how the monogamy rate can influence the spread of sexual transmitted disease in the
bi-partial network. By analyzing the basic reproductive number, we find that for the Poisson random network,
when the expectation number of neighbors is small, the the increase of monogamy rate can reduce
$R_0$. Meanwhile, for the scale free network, if the difference of neighbor number between different vertices
is big, then the increase of monogamy rate can reduce $R_0$. Then based on this model we do some simulations, and
the simulation results show that with the increase of the monogamy rate the total number of infected will decrease.
This work gives a new insight into how the monogamy rate can influence the transmission of sexual transmitted disease.
We believe if we could consider more human behaviors in our model, we could get more results about how different
human behaviors could influence the epidemics, thus gaining more ideas about how people can control or prevent the epidemics from occurring.

\bibliographystyle{plain}
\bibliography{ref}

\begin{thebibliography}{10}

\bibitem{dietz1993estimation}
Klaus Dietz.
\newblock The estimation of the basic reproduction number for infectious
  diseases.
\newblock {\em Statistical methods in medical research}, 2(1):23--41, 1993.

\bibitem{fox2001gonorrhea}
Kimberley~K Fox, Carlos Del~Rio, King~K Holmes, EW~Hook~3rd, Franklyn~N Judson,
  Joan~S Knapp, Gary~W Procop, Susan~A Wang, WL~Whittington, and William~C
  Levine.
\newblock Gonorrhea in the hiv era: a reversal in trends among men who have sex
  with men.
\newblock {\em American Journal of Public Health}, 91(6):959, 2001.

\bibitem{grant2010preexposure}
Robert~M Grant, Javier~R Lama, Peter~L Anderson, Vanessa McMahan, Albert~Y Liu,
  Lorena Vargas, Pedro Goicochea, Mart{\'\i}n Casap{\'\i}a, Juan~Vicente
  Guanira-Carranza, Maria~E Ramirez-Cardich, et~al.
\newblock Preexposure chemoprophylaxis for hiv prevention in men who have sex
  with men.
\newblock {\em New England Journal of Medicine}, 363(27):2587--2599, 2010.

\bibitem{grosskopf2014use}
Nicholas~A Grosskopf, Michael~T LeVasseur, and Debra~B Glaser.
\newblock Use of the internet and mobile-based “apps” for sex-seeking among
  men who have sex with men in new york city.
\newblock {\em American journal of men's health}, 8(6):510--520, 2014.

\bibitem{miller2011edge}
Joel~C Miller, Anja~C Slim, and Erik~M Volz.
\newblock Edge-based compartmental modelling for infectious disease spread.
\newblock {\em Journal of the Royal Society Interface}, 9(70):890--906, 2011.

\bibitem{muessig2013putting}
Kathryn~E Muessig, Emily~C Pike, Beth Fowler, Sara LeGrand, Jeffrey~T Parsons,
  Sheana~S Bull, Patrick~A Wilson, David~A Wohl, and Lisa~B Hightow-Weidman.
\newblock Putting prevention in their pockets: developing mobile phone-based
  hiv interventions for black men who have sex with men.
\newblock {\em AIDS patient care and STDs}, 27(4):211--222, 2013.

\bibitem{newman2002spread}
Mark~EJ Newman.
\newblock Spread of epidemic disease on networks.
\newblock {\em Physical review E}, 66(1):016128, 2002.

\bibitem{peng2013vaccination}
Xiao-Long Peng, Xin-Jian Xu, Xinchu Fu, and Tao Zhou.
\newblock Vaccination intervention on epidemic dynamics in networks.
\newblock {\em Physical Review E}, 87(2):022813, 2013.

\bibitem{ruan2012epidemic}
Zhongyuan Ruan, Ming Tang, and Zonghua Liu.
\newblock Epidemic spreading with information-driven vaccination.
\newblock {\em Physical Review E}, 86(3):036117, 2012.

\bibitem{volz2008sir}
Erik Volz.
\newblock Sir dynamics in random networks with heterogeneous connectivity.
\newblock {\em Journal of mathematical biology}, 56(3):293--310, 2008.

\bibitem{wikisamesex}
Wikipedia.
\newblock Same-sex marriage.
\newblock \url{https://en.wikipedia.org/wiki/Same-sex_marriage}.
\newblock Accessed Sep. 12th, 2019.

\bibitem{xiao2011modelling}
Yanni Xiao, Yicang Zhou, and Sanyi Tang.
\newblock Modelling disease spread in dispersal networks at two levels.
\newblock {\em Mathematical medicine and biology: a journal of the IMA},
  28(3):227--244, 2011.

\end{thebibliography}

\end{document}